\documentclass[twocolumn,showpacs,preprintnumbers,amsmath,amssymb]{revtex4}
\usepackage{graphicx}% Include figure files
\usepackage{dcolumn}% Align table columns on decimal point
\usepackage{amsfonts}
\usepackage{amsgen}
\usepackage{amsbsy}
\usepackage{bm}% bold math
\usepackage{hyperref}% makes the references hyperlinked
\newcommand{\hodge}[1]{\,*#1}
\newcommand{\lie}{{\cal L}}
\begin{document}
\title{$\mathcal{N}=1$ SUGRA Noether Charges}
\author{Rodrigo Aros}
\affiliation{Universidad Andr\'es Bello, Avenida Republica 237, Santiago,Chile}
\date{\today}
\begin{abstract}
In this work a generic set of boundary conditions for $\mathcal{N}=1$ SUGRA is proposed. This
conditions defines that Hamiltonian charges equals Noether ones, including supercharge.
\end{abstract}
\maketitle
\section{introduction}

The role of boundary conditions in any theory of fields is manyfold and fundamental. First the
boundary conditions are necessary to solve the equations of motion. They also are necessary to
have a proper action principle. Furthermore \textit{they define the ensemble in which the theory
exists} (for gravity see for instance  \cite{Brown:1993bq}).

In general the variation of the action, as a functional of a field $\phi(x)$, is given by
\[
 \delta I = \int_{\mathcal{M}}  EM \delta \phi + d\Theta(\phi,\delta\phi),
\]
where $EM$ stands for the equations of motion. Now the action is called stationary whether the
boundary term $\Theta(\phi,\delta\phi)$ vanishes on shell, which only happens provided adequate
boundary conditions. Only in this case one has a proper action principle.

For gravity the issue of an adequate set of boundary conditions has been certainly addressed by
many authors, in many different problems and approaches
(\cite{Abbott:1982ff,Teitelboim:1984kf,Henneaux:1985tv,Henneaux:1999ct,Barnich:2001jy}, etc. See
also \cite{Aros:1999id}).  In \cite{Aros:1999id,Aros:2001gz} for locally asymptotical AdS spaces
was introduced a new set of boundary condition at the asymptotical space region. As a result of
these new boundary conditions Noether and Hamiltonian charges for diffeomorphisms agree. The
intension of this work was extend the results in \cite{Aros:1999id,Aros:2001gz} to four
dimensional $\mathcal{N}=1$ supergravity (SUGRA). The results obtained are analogous.

To simplify the calculations only $\mathcal{N}=1$ SUGRA will be discussed, however most the
computations can be readily extended to $\mathcal{N}>1$ SUGRA's. This particular supergravity in
four dimensions is a simpler laboratory to test technics to later be applied in higher dimensions
or $\mathcal{N}>1$. In eleven dimensions, for instance, although there is no standard SUGRA with a
negative cosmological constant \cite{Bautier:1997yp} there is the eleven dimension Chern Simons
SUGRA \cite{Troncoso:1997me}.

The space to be discussed in this work is given by $\mathcal{M}=\mathbb{R}\times \Sigma$ where
$\Sigma$ corresponds to a 3-dimensional spacelike hypersurface and $\mathbb{R}$ stands for the
time direction. The space possesses an asymptotical locally AdS region, which defines the boundary
$\mathbb{R}\times\partial\Sigma_{\infty}$.

For simplicity in this work the differential forms language will be used, and $\wedge$ product
between differential forms will be omitted.

\section{Supergravity}
The four dimensional $\mathcal{N}=1$ supergravity action with negative cosmological constant is
defined by the Lagrangian \cite{Freund:1986ws}
\begin{equation}\label{SuperAction}
{\mathbf{L}} = \frac{l^{2}}{64\pi}\bar{R}^{ab}\bar{R}^{cd}\epsilon_{abcd}+
\bar{\Psi}\gamma_{5}\gamma_{a} e^{a} D \Psi,
\end{equation}
where $D=d + \frac{1}{4}\omega^{ab}\gamma_{ab}+ \frac{1}{2l}e^{a} \gamma_{a}$ and
$\bar{R}^{ab}=R^{ab}+ l^{-2}e^{a} e^{b}$. Here $e^{a}$ is the vierbein and $\omega^{ab}$ the spin
connection, $R^{ab}$ is the two form of curvature defined as
\[
R^{ab}=d\omega^{ab}+\omega^{a}_{\hspace{1ex} c}\omega^{cb}= \frac{1}{2}R^{ab}_{\hspace{2ex}cd}
\,e^{c} e^{d}
\]
where $R^{ab}_{\hspace{2ex}cd}$ is the Riemann Tensor. $\Psi$ stands for the Ravita Schwinger
field.

From the Lagrangian (\ref{SuperAction}) one obtains the equations of motion
\begin{subequations}
\label{Eq.Motion}
\begin{eqnarray}\label{Einstein}
\frac{l^{2}}{32\pi}\bar{R}^{ab} e^{c}\epsilon_{abcd} +  \bar{\Psi}\gamma_{5}\gamma_{d} D\Psi -
\frac{1}{2}\bar{\Psi}\gamma_{5}\gamma_{a}
  e^{a} \gamma_{d} \Psi&=0,\\
\label{Pre-Torsion}
 \frac{l^{2}}{32\pi}D( \bar{R}^{cd}\epsilon_{abcd})+\frac{1}{2}\bar{\Psi}\gamma_{5}\gamma_{c} e^{c}
 \gamma_{ab} \Psi&=0,
\end{eqnarray}
\begin{eqnarray}
\label{LeftSpinor}
    \gamma_{5}\gamma_{c} e^{c} D \Psi&=0,\\
\label{RightSpinor} D(\bar{\Psi} \gamma_{5}\gamma_{c}  e^{c})&=0.
\end{eqnarray}
\end{subequations}
and  the boundary term
\begin{equation}\label{BoundaryTerm}
\Theta = \frac{l^{2}}{32\pi}\left(\delta_0\omega^{ab} \bar{R}^{cd}\epsilon_{abcd} \right) +
\bar{\Psi}\gamma_{5}\gamma_{a}e^{a}\delta_0 \Psi.
\end{equation}

Remarkably Eq.(\ref{Pre-Torsion}) can be rewritten in terms of the torsion two form $T^{a}=de^{a}
+ \omega^{a}_{\hspace{1ex} b}e^{b}$ simply as
\begin{equation}\label{Torsion}
    T^{a} = - 4\pi \bar{\Psi} \gamma^{a} \Psi.
\end{equation}
The torsion two form contains the torsion tensor, $T^{a}_{bc}$, as $T^{a}=\frac{1}{2}T^{a}_{bc}
e^{b} e^{c}$ .

\section{Boundary Conditions}
In order to solve the equations of motion (\ref{Eq.Motion}) is obviously necessary to provide
boundary conditions. In Ref.\cite{Aros:1999id} a new set boundary conditions was introduced for
the bosonic sector of action (\ref{SuperAction}), namely the ALAdS condition. This boundary
conditions is based on that for any asymptotically locally AdS space one can define a pseudo
connection $W=1/2\, W^{AB}J_{AB}$, with
\[
 W^{AB} =  = \left[\begin{array}{cc}
   \omega^{ab} & l^{-1}e^a \\
   -l^{-1}e^a  & 0 \\
 \end{array} \right],
 \]
whose field strength $F=1/2\,F^{AB}J_{AB}= dA + A\wedge A$, with
\[
 F^{AB} = \left[\begin{array}{cc}
   \bar{R}^{ab} & l^{-1}T^a \\
   -l^{-1}T^a  & 0 \\
 \end{array} \right],
 \]
vanishes on the asymptotic spatial region. Since for the standard solutions $T^{a}=0$ everywhere,
the condition restricts only the curvature $R^{ab}$.

To incorporate the fermionic sector one option is to follow the same underlying idea, namely that
the (super)-field strength vanishes at asymptotical region. Analogously it is defined a
(super)-connection $A=1/2\,\omega^{ab} J_{ab} + l^{-1}e^a J_{a} +\bar{\Psi }Q$ whose field
strength reads
\begin{eqnarray}
F &=& \frac{1}{2} \left(\bar{R}^{ab} + \frac{1}{2} \bar{\Psi} \gamma^{ab} \Psi \right)J_{ab}
   + \left(\frac{T^{a}}{l} + \frac{1}{2}\bar{\Psi} \gamma^{a} \Psi \right)J_{a5} \nonumber\\
   &+& D\bar{\Psi } Q.\label{FieldStrength}
\end{eqnarray}

It is direct to see that to impose the vanishing of the field strength (\ref{FieldStrength}) at
the asymptotical region in turn implies that asymptotically $\bar{R}^{ab} \rightarrow -\frac{1}{2}
\bar{\Psi} \gamma^{ab} \Psi$, $T^{a} \rightarrow -\frac{1}{2}\bar{\Psi} \gamma^{a} \Psi$ and
$D\bar{\Psi}\rightarrow 0$. Unfortunately these boundary conditions do not imply the vanishing of
the boundary term (\ref{BoundaryTerm}). To overcome this one can also impose that $\Psi$ vanish at
the asymptotical region. This condition was determined in \cite{Henneaux:1985tv} in another
approach (See also \cite{Abbott:1982ff}). From the point of view of the gauge super group $\Psi$
is part of the connection, therefore to impose $\Psi=0$ at the asymptotical region can be regarded
as a gauge fixing.

The vanishing of $\Psi$ together with the vanishing of $F$ imply that at the asymptotic region
\[
\bar{R}^{ab} \rightarrow 0 \textrm{ and }T^{a}\rightarrow 0,
\]
namely that $\mathcal{M}$ is actually an asymptotically locally anti de Sitter space.

Now it is straightforward to prove that under these boundary conditions just proposed,
\textit{.i.e.},
\[F(x)\rightarrow 0 \textrm{ as } x \rightarrow \mathbb{R} \times \partial \Sigma_{\infty},\]
the boundary term (\ref{BoundaryTerm}) vanishes for arbitrary variations of the fields.

\section{Diffeomorphisms Noether current}
The invariance under diffeomorphisms of the Lagrangian (\ref{SuperAction}) yields a Noether
current (See (\ref{currentdensity}) in appendix) which, after a straightforward computation, reads
\begin{equation}\label{SuperNoetherCurrent}
 \hodge{{\bf J}_\xi} = -d \left(\frac{1}{32\pi}I_{\xi}\omega^{ab}\bar{R}^{cd}
 \epsilon_{abcd} + \bar{\Psi}\gamma_{5}\gamma_{a} e^{a}I_{\xi} \Psi \right).
\end{equation}

To define a conserved charge from the current in Eq.(\ref{SuperNoetherCurrent}) one must restrict
$\xi$ to be a global isometry, namely a Killing vector. On the other hand, since Eq.
(\ref{SuperNoetherCurrent}) is an exact form the conserved charge to be obtained form it, through
integration, depends only on its asymptotical value.

Note that the boundary condition at the asymptotical spatial region requires that $\Psi$ vanish
yielding the vanishing of the second term in Eq. (\ref{SuperNoetherCurrent}). The same result can
be obtained smoothly by proposing for the gravitino a fall-off behavior as $\Psi\sim r^{-3/2}$
(See Ref. \cite{Henneaux:1985tv}). This implies that any conserved charge associated with the
global diffeomorphisms reads
\[
Q_{\xi} = -\frac{1}{32\pi}\int_{\Sigma_{\infty}} I_{\xi}\omega^{ab}\bar{R}^{cd}
 \epsilon_{abcd}.
\]

Remarkably this result is the same obtained in Ref. \cite{Aros:1999id} for only the bosonic part
of Eq.(\ref{SuperAction}). This result indicates that the value of mass or angular momenta as the
asymptotical value of an integral should encode the presence of the gravitinos. This not usual,
for instance, in the case of Reissner Nordstr\"{o}m solution the expression of the charge associated
time symmetry reads
\[
 Q\left(\frac{\partial}{\partial t}\right) \sim  M - \frac{e^{2}}{r},
\]
which obviously only equals $M$, the mass, as $r\rightarrow \infty$.

This result also might let room for the detection of gravitinos as a correction to the value of
the mass at short distance.

\section{Supersymmetric Noether current}

It is well known that the Lagrangian (\ref{SuperAction}) is (on-shell) also invariant under the
supersymmetry transformation
\begin{equation}\label{SupertTransformation}
   \begin{array}{ll}
     \delta_{\epsilon}e^{a} &= 8\pi \bar{\epsilon}\gamma^{a}\Psi, \\
     \delta_{\epsilon}\Psi &= D\epsilon,\\
     \delta_{\epsilon}\omega^{ab}_{\,\,\,\mu} &= \pm\frac{1}{2}\epsilon^{abcd}
     \bar{\epsilon}\gamma_{5} \gamma_{c} E_{d}^{\,\,\nu} (D(\Psi))_{\mu\nu},
   \end{array}
\end{equation}
where $\epsilon$ is  a Grassmann valued $0$-form and $E_{a}^{\,\,\mu}$ is the inverse of
$e^{a}_{\,\mu}$.

After some manipulations it can be shown the Lagrangian (\ref{SuperAction}) changes under the
supersymmetric transformation (\ref{SupertTransformation}) in the boundary term
\[
\alpha=  \frac{l^{2}}{32\pi} (\delta_{\epsilon}\omega^{ab}\bar{R}^{cd}\varepsilon_{abcd} ) +
\bar{\epsilon} \gamma_{5}\gamma_{a} e^{a} D \Psi,
\]
and thus this supersymmetric transformation give rise to the Noether current $\hodge{{\bf
J}_\epsilon}=\Theta-\alpha$ which reads
\begin{equation}\label{SuperCharge}
\hodge{{\bf J}_\epsilon} = -d\left(\bar{\epsilon}\gamma_{5}\gamma_{a}e^{a}\Psi\right).
\end{equation}

Similarly one can define a charge associated with Eq.(\ref{SuperCharge}) by integrating at the
asymptotical spatial region as
\begin{equation}\label{SuperChargeReal}
Q_{\epsilon} = -\int_{\partial \Sigma_{\infty}}
\left(\bar{\epsilon}\gamma_{5}\gamma_{a}e^{a}\Psi\right).
\end{equation}
Remarkably, after some algebraic manipulations, Eq. (\ref{SuperChargeReal}) reproduces the
Hamiltonian prescription for the super charge obtained in \cite{Teitelboim:1977hc}.

\section{Hamiltonian versus Noetherian}

As far it has been found expressions for the Noether charges associated diffeomorphisms and
supersymmetry. However charges as mass and angular momenta or supercharges necessarily are defined
within the context of Hamiltonian formalism. To establish a connection between Noether and
Hamiltonian charges the covariant phase space method develop in \cite{Lee:1990nz} can be used.
For diffeomorphisms the variation of the Hamiltonian Charge $G_{\xi}$ reads
\begin{eqnarray}
   \delta G_{\xi} &=& \int_{\partial \Sigma} \delta \left(\frac{1}{32\pi}I_{\xi}\omega^{ab}\bar{R}^{cd}
 \epsilon_{abcd} + \bar{\Psi}\gamma_{5}\gamma_{a} e^{a}I_{\xi} \Psi \right)\nonumber \\
   &+ & I_{\xi} \Theta(\delta \omega, \delta \Psi, \omega, \Psi, e )\label{VarHamiltonian},
\end{eqnarray}
where $\Theta(\delta \omega, \delta \Psi, \omega, \Psi, e )$ is given by Eq.(\ref{BoundaryTerm}).
However the second term in Eq.(\ref{VarHamiltonian}) was tailored to vanish in the asymptotical
spatial region for arbitrary variations of the fields, thus
\begin{equation}\label{VarHamiltonia-Noether}
 \delta G_{\xi} = \delta Q_{\xi}.
\end{equation}

In this way the result obtained in Ref. \cite{Aros:2001gz} for the bosonic part of Eq.
(\ref{SuperAction}) extends to SUGRA with $\mathcal{N}=1$. Furthermore note that this result
allows to conjecture, given the structure of Eq.(\ref{VarHamiltonian}), that this result could be
extended to others SUGRA with negative cosmological constant in higher dimensions and $\mathcal{N}
\neq 1$.

Similarly the expression of the Noether charge associated with supersymmetry can be proven to be
equivalent to the Hamiltonian one, \textit{i.e.}
\begin{equation}\label{VariationofSuperCurrent}
    \delta G_{\epsilon} = \delta Q_{\epsilon}.
\end{equation}
This results is not surprising since Eq.(\ref{SuperChargeReal}) corresponds to the Hamiltonian
prescription.

\section{Conclusions}

In this work were analyzed the Noether currents associated with symmetry of diffeomorphisms and
the supersymmetry of the $\mathcal{N}=1$ supergravity in four dimensions with a negative
cosmological constant. It was proven that it is possible to define a proper set of boundary
conditions which not only defines a sensible action principle, but also determines that Noether
charges equal Hamiltonian ones. It is wroth to mention that the results obtained in this work are
equivalent to those in \cite{Henneaux:1999ct} in terms of superpotentials.

The boundary conditions proposed in this paper are not new but corresponds to a combination of an
extension of previous results of the bosonic part of the action and previously known prescription
for the asymptotical behavior of gravitinos.

The generic form in which the results arise allows to conjecture that the basic results of this
work should extend to other SUGRA in higher dimensions.

\begin{acknowledgments}
This work is partially funded by grants FONDECYT 1040202 and DI 06-04. (UNAB).
\end{acknowledgments}

\appendix

\section{Noether Method}

Any infinitesimal transformation of a field $\phi(x)$ can be split as a local functional
transformation plus a diffeomorphism, given by $x' = x + \xi(x)$ (See \cite{Ramond:1989yd}). This
can be done as follows,
\begin{equation}\label{SplitingTheTransf}
\delta \phi(x)= \phi'(x') - \phi(x)= \phi'(x') - \phi'(x) + \phi'(x) - \phi(x)
\end{equation}
where $\phi'(x) - \phi(x)= \delta_{0} \phi(x)$ is a local functional transformation, and
$\phi'(x') - \phi'(x) = {\cal L}_\xi \phi$ corresponds to the Lie derivative a along the vector
field $\xi$, thus $\delta\phi = \delta_{0}\phi + {\cal L}_\xi \phi$. For any Lagrangian this
yields
\begin{equation}\label{LagrangianVariation}
  \delta {\mathbf{L}} = (E.M.)\delta_{0} \phi + d \Theta(\phi,\delta_{0}\phi) + d I_\xi
  {\mathbf{L}}.
\end{equation}

A symmetry is defined as a change in the field configuration that does not alter the field
equations. To satisfy that $\delta\phi$ has to be such that the Lagrangian changes in a total
derivative, i.e., $\delta {\mathbf{L}} = d \alpha$, thus on shell the current
\begin{equation}\label{currentdensity}
 \hodge{{\bf J}_\xi} = \Theta(\delta_{0} \phi,\phi) + I_\xi{\bf L}-\alpha,
\end{equation}
satisfies $d(\hodge{{\bf J}_\xi})=0$, which is usually called the Noether current.

\section{Defining the AdS Algebra}

In this work is used the algebra definition
\begin{eqnarray*}
  [J_{AB},J_{CD}] &=& -\delta^{EF}_{AB}\delta^{GH}_{CD}\eta_{EG}J_{FH} \\
  \left[Q^{\alpha},J_{CD}\right] &=& (\gamma_{CD})^{\alpha}_{\,\,\beta} Q^{\beta} \\
  \{Q^{\alpha},Q^{\beta}\} &=& \frac{1}{2}\left(\gamma^{AB} C^{-1}\right)^{\alpha\beta} J_{AB}
\end{eqnarray*}

Additionally since the Clifford algebra is realized by the $\gamma $ matrices as $\left\{ \gamma
_{a},\gamma _{b}\right\} =2\eta _{ab}$, the AdS generators have the representation given by
\[
 J_{ab} =\frac{1}{2}\gamma _{ab}  \textrm{ and } J_{a5} =\frac{1}{2}\gamma _{a}
\]
with $\gamma _{ab}=\frac{1}{2}[\gamma _{a},\gamma _{b}]$.

In addition the representation satisfies that,
\begin{eqnarray*}
 C^{T} &=& - C\\
 C \gamma_{a} C^{-1} &=& -\gamma_{a}^{T}.
\end{eqnarray*}

\section{Useful relations}
During the calculations in this work the useful identity of the commutator the covariant
derivative $D= d + \frac{1}{4}\omega^{ab}\gamma_{ab}+ \frac{1}{2l}e^{a} \gamma_{a}$ and the lie
derivative, which reads,
\begin{equation}
    [\lie_{\xi},D]\Psi = \left(\lie_{\xi}\omega^{ab}\right)J_{ab} \Psi=
    \left(D(I_{\xi}\omega^{ab})+ I_{\xi}R^{ab}\right)\Psi,
\end{equation}
was necessary.

In addition these other identities
\begin{eqnarray*}
\gamma_{5}&=&\gamma_{0}\gamma_{1}\gamma_{2}\gamma_{3}\\
   \epsilon^{abcd}\gamma_{cd} &=& 2\gamma_{5}\gamma^{ab}\\
   \gamma_{ab} \gamma_{c} &=&
   (\eta_{bc}\gamma_{a}-\eta_{ac}\gamma_{b})+\epsilon_{abcd}\gamma_{5}\gamma^{d}\\
    \gamma_{c} \gamma_{ab} &=&
   (\eta_{ca}\gamma_{b}-\eta_{ab}\gamma_{a})+\epsilon_{abcd}\gamma_{5}\gamma^{d}\\
   \gamma_{a}\gamma_{cd}\gamma^{a}&=&0 \\\gamma_{cd}\gamma_{a}\gamma^{cd}&=&0
\end{eqnarray*}
for the gamma matrices were necessary as well.

%\bibliographystyle{jhep}
%\bibliography{myXbib}

\begin{thebibliography}{10}

\bibitem{Brown:1993bq}
J.~D. Brown and J.~W. York, {\it The microcanonical functional integral. 1. the
  gravitational field},  {\em Phys. Rev.} {\bf D47} (1993) 1420--1431,
  [\href{http://xxx.lanl.gov/abs/http://arXiv.org/abs/gr-qc/9209014}{{\tt
  http://arXiv.org/abs/gr-qc/9209014}}].

\bibitem{Abbott:1982ff}
L.~F. Abbott and S.~Deser, {\it Stability of gravity with a cosmological
  constant},  {\em Nucl. Phys.} {\bf B195} (1982) 76.

\bibitem{Teitelboim:1984kf}
C.~Teitelboim, {\it Manifestly positive energy expression in classical gravity:
  Simplified derivation from supergravity},  {\em Phys. Rev.} {\bf D29} (1984)
  2763--2765.

\bibitem{Henneaux:1985tv}
M.~Henneaux and C.~Teitelboim, {\it Asymptotically anti-de sitter spaces},
  {\em Commun. Math. Phys.} {\bf 98} (1985) 391--424.

\bibitem{Henneaux:1999ct}
M.~Henneaux, B.~Julia, and S.~Silva, {\it Noether superpotentials in
  supergravities},  {\em Nucl. Phys.} {\bf B563} (1999) 448--460,
  [\href{http://xxx.lanl.gov/abs/hep-th/9904003}{{\tt hep-th/9904003}}].

\bibitem{Barnich:2001jy}
G.~Barnich and F.~Brandt, {\it Covariant theory of asymptotic symmetries,
  conservation laws and central charges},  {\em Nucl. Phys.} {\bf B633} (2002)
  3--82,
  [\href{http://xxx.lanl.gov/abs/http://arXiv.org/abs/hep-th/0111246}{{\tt
  http://arXiv.org/abs/hep-th/0111246}}].

\bibitem{Aros:1999id}
R.~Aros, M.~Contreras, R.~Olea, R.~Troncoso, and J.~Zanelli, {\it Conserved
  charges for gravity with locally ads asymptotics},  {\em Phys. Rev. Lett.}
  {\bf 84} (2000) 1647, [\href{http://xxx.lanl.gov/abs/gr-qc/9909015}{{\tt
  gr-qc/9909015}}].

\bibitem{Aros:2001gz}
R.~Aros, {\it Analyzing charges in even dimensions},  {\em Class. Quant. Grav.}
  {\bf 18} (2001) 5359--5369,
  [\href{http://xxx.lanl.gov/abs/gr-qc/0011009}{{\tt gr-qc/0011009}}].

\bibitem{Bautier:1997yp}
K.~Bautier, S.~Deser, M.~Henneaux, and D.~Seminara, {\it No cosmological d = 11
  supergravity},  {\em Phys. Lett.} {\bf B406} (1997) 49--53,
  [\href{http://xxx.lanl.gov/abs/hep-th/9704131}{{\tt hep-th/9704131}}].

\bibitem{Troncoso:1997me}
R.~Troncoso and J.~Zanelli, {\it Chern-simons supergravities with off-shell
  local superalgebras},
  \href{http://xxx.lanl.gov/abs/http://arXiv.org/abs/hep-th/9902003}{{\tt
  http://arXiv.org/abs/hep-th/9902003}}.

\bibitem{Freund:1986ws}
P.~G.~O. Freund, {\em INTRODUCTION TO SUPERSYMMETRY}.
\newblock CAMBRIDGE MONOGRAPHS ON MATHEMATICAL PHYSICS. Cambridge Univ. Pr.,
  Cambridge U.K., 1986.
\newblock 152 P.

\bibitem{Teitelboim:1977hc}
C.~Teitelboim, {\it Surface integrals as symmetry generators in supergravity
  theory},  {\em Phys. Lett.} {\bf B69} (1977) 240--244.

\bibitem{Lee:1990nz}
J.~Lee and R.~M. Wald, {\it Local symmetries and constraints},  {\em J. Math.
  Phys.} {\bf 31} (1990) 725.

\bibitem{Ramond:1989yd}
P.~Ramond, {\em FIELD THEORY: A MODERN PRIMER,FRONTIERS IN PHYSICS, 74}.
\newblock ADDISON-WESLEY, REDWOOD CITY, USA, 1989.

\end{thebibliography}

\providecommand{\href}[2]{#2}\begingroup\raggedright\endgroup

\end{document}